\documentclass[showpacs,twocolumn,prb]{revtex4-1}
\usepackage{amsfonts}
\usepackage{graphicx}
\usepackage{amsmath}
\usepackage{amssymb}

\begin{document}

\title{Conductivity of Dirac fermions with phonon induced topological crossover}
\author{Zhou Li$^1$}
\email{lizhou@mcmaster.ca}
\author{J. P. Carbotte$^{1,2}$}

\affiliation{$^1$ Department of Physics, McMaster University,
Hamilton, Ontario,
Canada L8S 4M1 \\
$^2$ Canadian Institute for Advanced Research, Toronto, Ontario,
Canada M5G 1Z8}

\begin{abstract}
We study the Hall conductivity in single layer gapped Dirac fermion
materials including coupling to a phonon field, which not only
modifies the quasi-particle dynamics through the usual self-energy
term but also renormalizes directly the gap. Consequently the Berry
curvature is modified. As the temperature is increased the sign of
the renormalized gap can change and the material can cross over from
a band insulator to a topological insulator at higher temperature
(T). The effective Chern numbers defined for valley and spin Hall
conductivity show a rich phase diagram with increasing temperature.
While the spin and valley DC Hall conductivity is no longer
quantized at elevated temperature a change in sign with increasing T
is a clear indication of a topological crossover. The chirality of
the circularly polarized light which is dominantly absorbed by a
particular valley can change with temperature as a result of a
topological crossover.
\end{abstract}

\date{\today}
\maketitle

\section{Introduction}

The isolation of graphene by exfoliation \cite{Novo1,Novo2} led to
the discovery of many of its exotic properties \cite{Zhang, Geim}.
It also led to the fabrication and study of other two dimensional
systems such as single layer group-VI dichalcogenides
\cite{Chei,Zhu,Feng,Li} (e.g. $MoS_{2}$) and silicene
\cite{Aufray,Padova,Stille}, a buckled honeycomb lattice of silicon
atoms. The Kane-Mele \cite{Kane} low energy Hamiltonian can be used
to describe these materials which have Dirac cones and valley
degeneracy. The Dirac fermions can acquire mass and spin-orbit
coupling can spin polarize the bands. Manipulation of the size of
the gaps involved can lead to a topological phase transition from a
quantum spin Hall state to a band insulator. It is known from
consideration of the selection rules \cite{Xiao,Xiao1} for the
optical transition involving circularly polarized light that the two
valleys with indices $\tau=\pm 1$ react oppositely to the incident
photon. Each valley has opposite sign of orbital magnetic moment and
Berry curvature, and right (left) handed polarized radiation excites
preferentially valley $\tau=+1(-1)$. These materials display a rich
phase diagram with phases having different topological quantum
numbers (Chern numbers) and are believed to be ideal for
valleytronics \cite{Zeng,Mak,Cao}.

Here we go beyond a bare band model and include in the Hamiltonian
the coupling of the charge carriers to a phonon field
\cite{Li1,Carbotte,Stauber,Pound}. We find that the spin and valley
dependent gap is renormalized by the phonons and that this
renormalization depends strongly on temperature \cite{Ion}.
Consequently the Berry curvature will be modified and the
topological quantum numbers are not well protected. For example, as
we increase temperature, the gap can close and reopen and as found
by Garate \cite{Ion} the system can crossover from a band to a
topological insulator. While reference [22] was concerned only with
a proof in principle, here we show how this crossover can be
measured experimentally either using spin and/or valley DC Hall
conductivity or through the circular dichroism of absorbed light.
The Kane-Mele model Hamiltonian is specified in section II along
with electron-phonon coupling based on the Holstein model. The self
energy due to electron-phonon interaction directly modifies the gap
which changes sign with increasing temperature (T). Formulas for AC
longitudinal and transverse optical conductivity including self
energy effects are presented in section III. The change in sign of
the renormalized gap with increasing T can be viewed as changing the
underlying Chern number and we describe a pathway to determining
this topological crossover through DC valley and spin Hall effects.
Section IV deals with the temperature dependence of the dichroism of
circularly polarized light with the handedness of the dominant
absorption changing sign with increasing temperature. Section V
provides a brief conclusion.

\section{Formalism and renormalized gap}

The Kane-Mele model Hamiltonian \cite{Kane} for gapped Dirac
fermions with spin-orbit coupling takes the form
\begin{equation}
\widehat{H}_{0}=\hbar v(\tau k_{x}\hat{\sigma}_{x}+k_{y}\hat{\sigma}
_{y})+\Delta _{z}\hat{\sigma}_{z}-\lambda _{SO}\tau
\hat{\sigma}_{z}\hat{S}_{z}\text{,}  \label{H0}
\end{equation}
where the valley index $\tau =\pm 1$, $\lambda _{SO}$ is the spin
orbit coupling parameter, $\Delta _{z}$ a gap and $v$ is the Fermi
velocity. For definiteness in our calculations we will use $\lambda
_{SO}=4meV$ and $v\approx 5\times 10^{5}m/s$
\cite{Aufray,Padova,Stille}. The $\hat{\sigma}'s$ are the Pauli
matrices for pseudospin and $\hat{S}_{z}$ is the spin matrix for its
z-component ($s_{z}$). Here we define the spin and valley dependent
gap as $\Delta _{s_{z}}^{\tau }=\Delta _{z}-\lambda _{SO}\tau
s_{z}$.
\begin{figure}[tp]
\begin{center}
\includegraphics[height=5in,width=3in]{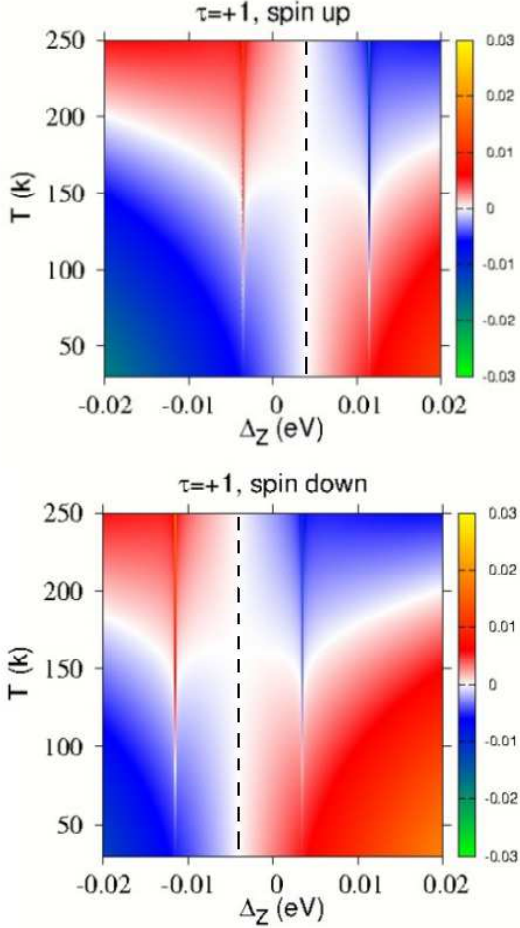}
\end{center}
\caption{(Color online) Temperature dependent renormalized gap
$\tilde{\Delta}_{s_{z}}^{\tau }$ for spin up (top frame) and spin
down (bottom frame) charge carriers, as a function of the bare mass
gap $\Delta _{z}$. The vertical black dashed line indicate the line
$\tilde{\Delta}_{s_{z}}^{\tau }=\Delta _{s_{z}}^{\tau }=Re\Sigma
^{Z}(\tau ,s_{z},0)=0$.} \label{fig1}
\end{figure}

\begin{figure}[tp]
\begin{center}
\includegraphics[height=3.5in,width=3.5in]{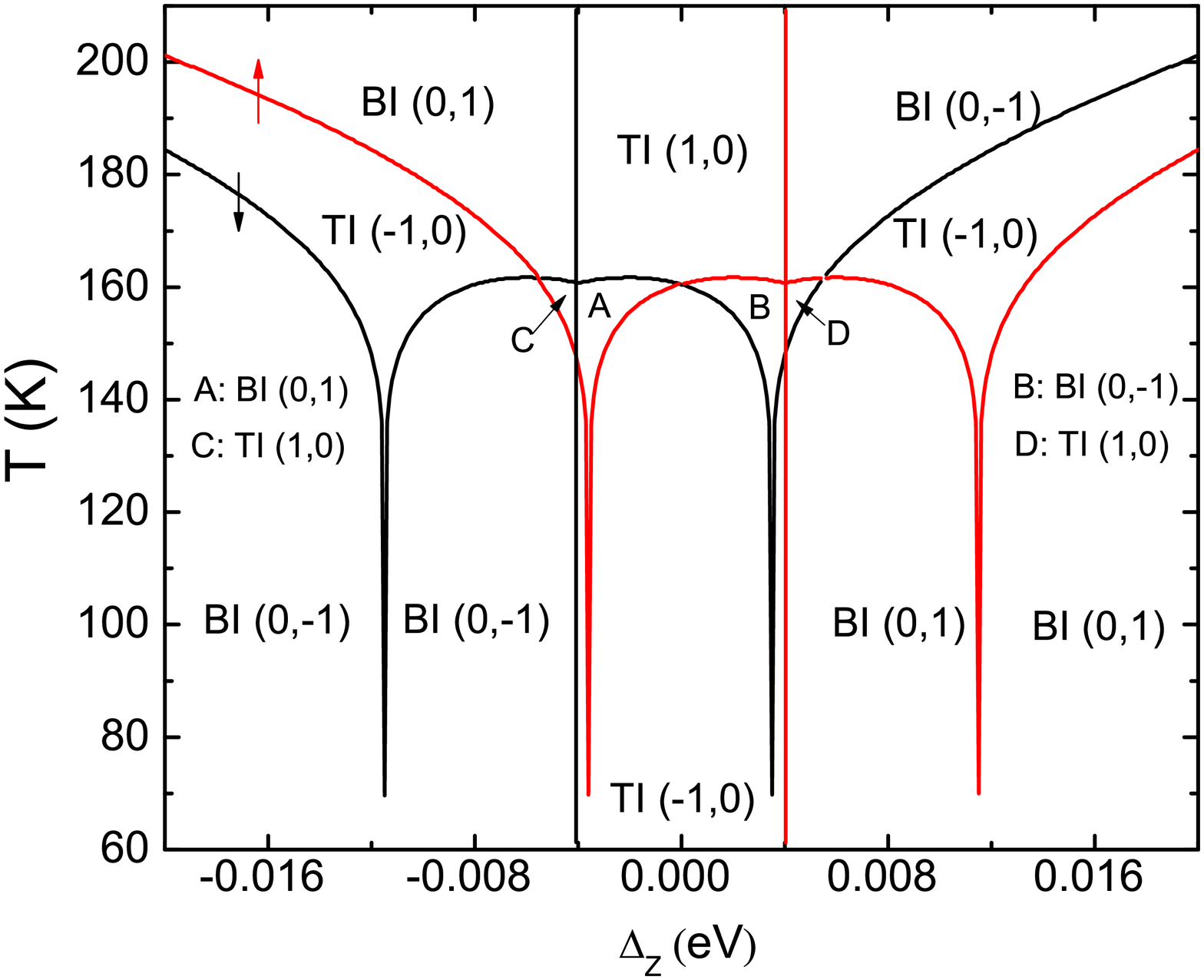}
\end{center}
\caption{(Color online) The phase diagram in the $\Delta_{z}$-$T$
plane.  The red and black line (phase boundaries) indicate
$\tilde{\Delta}_{s_{z}}^{\tau }=0$ for charge carriers in one valley
with spin up and down respectively. The notation is BI (x,y) and TI
(x,y) for band and topological insulator respectively with spin
Chern number x and valley Chern number y. Increasing temperature can
induce crossover from a band insulator to a topological insulator or
vice versa. } \label{fig2}
\end{figure}

\begin{figure}[tp]
\begin{center}
\includegraphics[height=4in,width=3.5in]{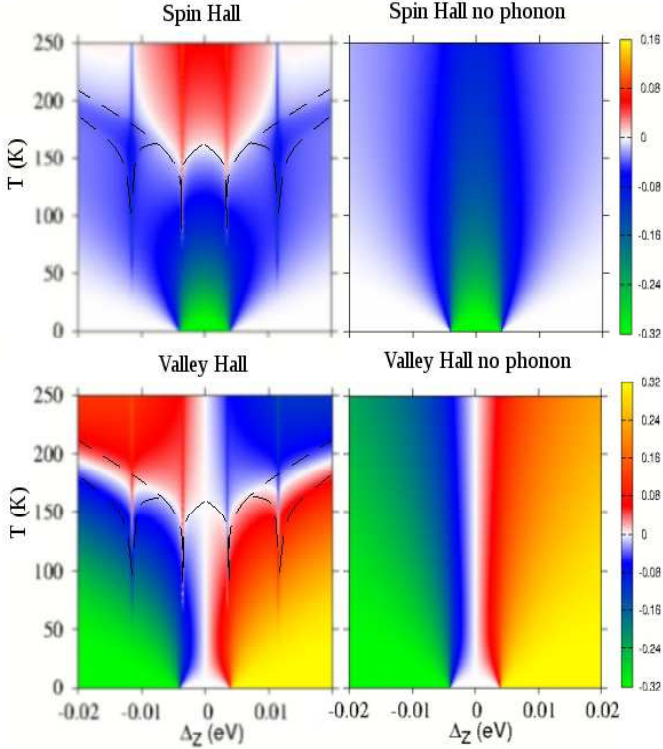}
\end{center}
\caption{(Color online) Color plot of the temperature dependent spin
(top) and valley (bottom) Hall conductivity as a function of
temperature $T$ and mass gap $\Delta_{z}$. Superimposed as long
dashed black lines are the phase boundaries of Fig.~\ref{fig2}.}
\label{fig3}
\end{figure}
The Holstein electron-phonon interaction which has been widely used,
is written as
\begin{equation}
H_{e-ph}=-g\omega _{E}\sum_{\mathbf{k},\mathbf{k}^{\prime },s}c_{\mathbf{k}%
,s}^{\dag }c_{\mathbf{k}^{\prime },s}(b_{\mathbf{k}^{\prime }-\mathbf{k}%
}^{\dag }+b_{\mathbf{k}-\mathbf{k}^{\prime }})+\sum_{\mathbf{q}}\omega
_{E}b_{\mathbf{q}}^{\dag }b_{\mathbf{q}}\text{,}  \label{phonon}
\end{equation}%
where $b_{\mathbf{q}}^{\dag }$ creates a phonon of energy $\omega
_{E}$ and momentum $\mathbf{q}$, $g$ is a coupling between electrons
and phonons and $c_{\mathbf{k},s}^{\dag }$ creates an electron of
momentum $\mathbf{k}$ and spin $s$. The self energy is given by
\cite{Li1}
\begin{equation}
\hat{\Sigma}\left( i\omega _{n}\right) =\Sigma ^{I}(i\omega _{n})\hat{I}%
+\Sigma ^{Z}(i\omega _{n})\hat{\sigma}_{z}
\end{equation}%
We are interested in the case when the chemical potential $\mu =0$,
so that $\mu$ lies in the gap. In this case the self energy
$Re\Sigma ^{I}(0)=0$ so the chemical potential will not be shifted
by the electron-phonon interaction. However $\Sigma ^{Z}(i\omega
_{n}\rightarrow \omega +i0^{+})$ will modify the gap $\Delta
_{s_{z}}^{\tau }$ at $\omega =0$ and consequently the Berry
curvature. We give an analytical expression here for the $\omega =0$
limit
\begin{eqnarray}
&&Re\Sigma ^{Z}(\tau ,s_{z},0)=\frac{g^{2}\omega _{E}^{2}\Delta
_{s_{z}}^{\tau }}{2t^{2}\pi }\times  \notag \\
&&[\ln |\frac{\omega _{E}+|\Delta _{s_{z}}^{\tau }|}{\omega
_{E}+\sqrt{x_{cut}}}|+\frac{\ln |(|\Delta _{s_{z}}^{\tau
}|^{2}-\omega_{E}^{2})/(x_{cut}^{2}-\omega _{E}^{2})|}{\exp (\omega
_{E}/(k_{B}T))-1}] \text{.} \label{self}
\end{eqnarray}%
Here $\sqrt{x_{cut}}=\sqrt{\hbar ^{2}v^{2}k_{\max }^{2}+(\Delta
_{s_{z}}^{\tau })^{2}}$ and $\hbar vk_{\max }=3.5eV$. The
electron-phonon coupling $g$ is chosen to make the effective mass
correction to be about $0.2$, which is a conservative estimate (a
recent experiment in $Cu_{x}Bi_2Se_3$ \cite{Kondo} found an order of
magnitude larger effective mass than what we have used here) of a
typical phonon coupling in semiconductors \cite{Hatch}, $\omega
_{E}$ is chosen to be $7.5meV$ and $t=\hbar v/a$ where $a$ is the
lattice constant, from this we know $t\approx1.0eV$.

For this set of parameters the first term in Eq.~(\ref{self}) which
gives the zero temperature value of the renormalization $Re\Sigma
^{Z}(\tau ,s_{z},0)$ provides a $22\%$ correction to $\Delta
_{s_{z}}^{\tau }$ which is small. The second term in
Eq.~(\ref{self}) however can become large as temperature increases
and $k_{B}T>>\omega _{E}$. The interacting Green's function is given
by
\begin{equation}
\hat{G}(\mathbf{k},i\omega _{n})=\frac{1}{2}\sum_{s=\pm
}(1+s\mathbf{H}_{\mathbf{k}}\cdot \mathbf{\sigma
})G(\mathbf{k},s,i\omega _{n})  \label{full}
\end{equation}
with
\begin{equation}
\mathbf{H}_{\mathbf{k}}=\frac{(\hbar v\tau k_{x},\hbar vk_{y},\Delta
_{s_{z}}^{\tau }+\Sigma ^{Z\ast }(i\omega _{n}))}{\sqrt{\hbar
^{2}v^{2}k^{2}+|\Delta _{s_{z}}^{\tau }+\Sigma ^{Z}(i\omega _{n})|^{2}}}
\label{Hk}
\end{equation}
and
\begin{equation}
G(\mathbf{k},s,i\omega _{n})=\frac{1}{i\omega _{n}+\mu -\lambda \tau
s_{z}/2-E(\mathbf{k},s,i\omega _{n})}  \label{Green}
\end{equation}
where $E(\mathbf{k},s,i\omega _{n})=s\sqrt{|\Delta _{s_{z}}^{\tau
}+\Sigma ^{z}(i\omega _{n})|^{2}+\hbar ^{2}v^{2}k^{2}}+\Sigma
^{I}(i\omega _{n})$. The self energy $\Sigma ^{I}(i\omega _{n})$
\cite{Carbotte,Stauber,Pound} directly renormalizes the
quasiparticle energies when there is no gap ($\Delta _{s_{z}}^{\tau
}=0$). However when $\Delta _{s_{z}}^{\tau }$ is non zero we see
that $\Sigma ^{Z}(i\omega _{n})$ enters Eq.~(\ref{Hk}) and
Eq.~(\ref{Green}) to directly renormalize the gap ($\Delta
_{s_{z}}^{\tau }$). Thus the Berry curvature is renormalized by the
electron-phonon interaction through the nontrivial gap self energy
$\Sigma ^{Z}(i\omega _{n})$. At zero temperature, only the $\omega
=0 $ limit of $\Sigma ^{Z}(i\omega _{n}\rightarrow \omega +i0^{+})$
enters the DC limit of the Hall conductivity which depends on the
renormalized gap $\tilde{\Delta}_{s_{z}}^{\tau }=\Delta
_{s_{z}}^{\tau }+Re\Sigma ^{Z}(\tau ,s_{z},0)$ and this plays a
critical role in this work. In Fig.~\ref{fig1} we show a color plot
of the magnitude of $\tilde{\Delta}_{s_{z}}^{\tau }$ as a function
of the bare gap $\Delta _{z}$ and temperature $T$ for fixed value of
$\lambda _{SO}$. The top frame is for one valley, $\tau =+1$, and
spin up, while the bottom frame is for spin down. The vertical
dashed line at
$\Delta_{z}=4$ and $-4meV$ respectively indicates the line $\tilde{\Delta}%
_{s_{z}}^{\tau }=\Delta _{s_{z}}^{\tau }=Re\Sigma ^{Z}(\tau
,s_{z},0)=0$. The renormalized gap has other zeros as indicated by
the white curves which separate positive and negative regions of
$\tilde{\Delta}_{s_{z}}^{\tau }$. This change in the sign of the gap
indicate possible topological crossovers as we will discuss below.

\section{Formalism and AC longitudinal and transverse conductivity}

The AC longitudinal conductivity $\sigma _{xx}(\omega )$ and
transverse Hall conductivity $\sigma _{xy}(\omega )$ in the lowest
order approximation, following from the Kubo formula without vertex
corrections \cite{Capp} are given by \cite{Li,Li1}
\begin{equation}
\sigma _{xx}(\omega )=\frac{e^{2}}{\hbar ^{2}}\int_{1,2}\sum_{\mathbf{k}%
}Tr\langle \sigma _{x}\hat{A}(\mathbf{k,}\omega _{1})\sigma _{x}\hat{A}(%
\mathbf{k,}\omega _{2})\rangle
\end{equation}
where we are only interested in the absorptive part of the
conductivity so no diamagnetic part is required \cite{Benf},
\begin{equation}
\sigma _{xy}(\omega )=\frac{e^{2}}{\hbar ^{2}}\int_{1,2}\sum_{\mathbf{k}%
}Tr\langle \tau _{z}\sigma _{x}\hat{A}(\mathbf{k,}\omega _{1})\sigma _{y}%
\hat{A}(\mathbf{k,}\omega _{2})\rangle\label{cxy}
\end{equation}
where $\hat{A}(\mathbf{k},\omega )$ is the matrix spectral density,
$\int_{1,2}=\int_{-\infty }^{\infty }d\omega _{1}d\omega
_{2}F(\omega)$ and the function $F(\omega )=-\frac{\hbar ^{2}v^{2}}{4\pi ^{2}i\omega }%
\frac{[f(\omega _{1})-f(\omega _{2})]}{\omega -\omega _{2}+\omega
_{1}+i\delta }$. The optical conductivity for circularly polarized
light follows as
\begin{equation}
Re\sigma _{\pm }(\omega )=Re\sigma _{xx}(\omega )\mp Im\sigma
_{xy}(\omega )
\end{equation}
Details on the impact of the self energies $\Sigma ^{I}(i\omega
_{n})$ and $\Sigma ^{Z}(i\omega _{n})$ on the longitudinal and Hall
optical conductivity are found in the reference \cite{Li1} to which
the reader is referred.
\begin{figure}[tp]
\begin{center}
\includegraphics[height=3.5in,width=3.5in]{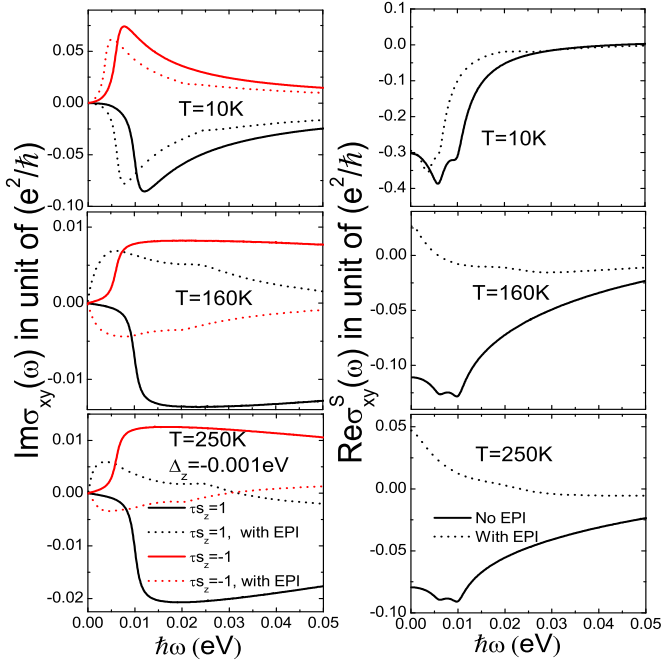}
\end{center}
\caption{(Color online) The real (right frames) and imaginary (left
frames) of the transverse conductivity $\sigma_{xy}(\omega)$ in
units of $e^2/\hbar$ vs. photon energy in eV for three temperatures
$T=10K$, $160K$ and $250K$ from top to bottom. In all cases the gap
$\Delta_z=0.001eV$. Black lines are for $\tau s_z=1$ and red for
$\tau s_z=-1$. Solid curves are bare band results while dotted
curves include the electron-phonon renormalizations and are based on
Eq.~(\ref{cxy}).} \label{fig4}
\end{figure}

\begin{figure}[tp]
\begin{center}
\includegraphics[height=3.5in,width=3.5in]{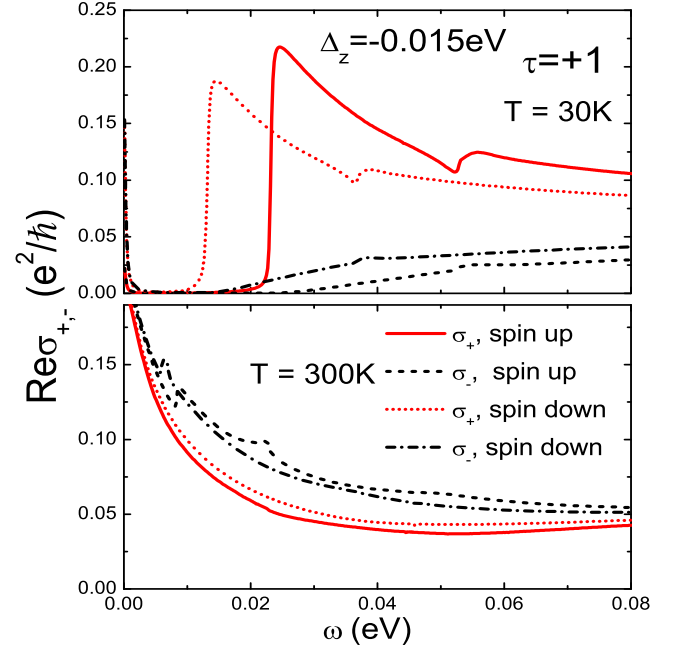}
\end{center}
\caption{(Color online) Temperature dependence of the dynamic
optical conductivity of the circularly polarized light. The top
frame is for $T=30K$ (low temperature) and the bottom frame for
$T=300K$ (high temperature). The valley index is $\tau=1$. Red lines
are for right hand polarized light and the black for left hand. The
contribution from spin up and spin down bands are given separately.}
\label{fig5}
\end{figure}

We begin our discussion with the zero temperature DC limit of the
spin and valley Hall conductivity, which can be associated to the
spin and valley Chern numbers. Finite temperature will obscure this
relationship, but we will still speak of an effective Chern number
using the zero frequency renormalized gap calculated at finite $T$
in the zero temperature bare band expressions for the
conductivities. At finite temperature the Hall conductivity is then
given by
\begin{equation}
Re\sigma _{xy}=\frac{e^{2}}{4\pi \hbar }\sum_{\tau ,s_{z}}\tau \int kdk\frac{%
\hbar ^{2}v^{2}\tilde{\Delta}_{s_{z}}^{\tau }[f(\varepsilon _{\mathbf{k}%
,-})-f(\varepsilon _{\mathbf{k},+})]}{[|\tilde{\Delta}_{s_{z}}^{\tau
}|^{2}+\hbar ^{2}v^{2}k^{2}]^{3/2}}  \label{xy}
\end{equation}%
with $f(\omega)$ the Fermi-Dirac distribution function. For the spin
and valley Hall conductivity $\tau $ in Eq.~(\ref{xy}) should be
replaced by $\hbar \tau s_{z}/(2e)$ and $1/e$ respectively. At $T=0$
Eq.~(\ref{xy}) reduces to
\begin{equation}
Re\sigma _{xy}=\frac{e^{2}}{4\pi \hbar }\sum_{\tau ,s_{z}}\tau \lbrack sgn(%
\tilde{\Delta}_{s_{z}}^{\tau })]\label{xy0}
\end{equation}
thus the spin and valley Hall conductivity remains quantized at T=0
in our treatment of the electron-phonon interaction. Its only effect
is to change the value of $\tilde{\Delta}_{s_{z}}^{\tau }$ over its
bare band value through the renormalization $Re\Sigma ^{Z}(\tau
,s_{z},0)$ of Eq.~(\ref{self}). The Chern number, spin Chern number
and valley Chern number are defined as $C=\frac{1}{4}\sum_{\tau
,s_{z}}\tau \lbrack sgn(\tilde{\Delta}_{s_{z}}^{\tau
})]$, $C_{s}=\frac{1}{4}\sum_{\tau ,s_{z}}\tau s_{z}[sgn(\tilde{\Delta}%
_{s_{z}}^{\tau })]$ and $C_{v}=\frac{1}{4}\sum_{\tau ,s_{z}}[sgn(\tilde{%
\Delta}_{s_{z}}^{\tau })]$. A phase diagram for the effective Chern
numbers obtained in this way is given in Fig.~\ref{fig2} with
$\Delta _{z}$ on the horizontal axis and temperature $T$ on the
vertical axis. The solid black lines give the zeros of the
renormalized gap $\tilde{\Delta}_{s_{z}}^{\tau }\equiv \Delta
_{s_{z}}^{\tau }+Re\Sigma ^{Z}(\tau ,s_{z},0)$. The labels TI and BI
indicate topological and band insulator respectively with right hand
integer $0,\pm 1$ the effective valley Chern number and the left
hand integer the effective spin Chern number. Once again we
emphasize that these numbers are obtained from a $T=0$ calculation
for $\sigma _{xy}$ of Eq.~(\ref{xy0}) using the value of the
renormalized gap obtained at finite $T$. Of course for finite $T$
one should really calculate the valley and spin Hall conductivity at
the same $T$ using the complete Eq.~(\ref{cxy}). Nevertheless
Fig.~\ref{fig2} remains useful in defining the various regions
involved and we will see that the phase boundaries defined in this
figure remain imprinted in our finite $T$ results although these
boundaries are definitely fuzzed out by temperature. This is shown
in Fig.~\ref{fig3} where we give a color plot of the spin and valley
Hall DC conductivity obtained from the finite temperature formula
Eq.~(\ref{xy}). This equation is a simplification of Eq.~(\ref{cxy})
for the Hall conductivity. Eq.~(\ref{cxy}) includes the complete
frequency dependent renormalization due to both $\Sigma ^{I}(i\omega
_{n})$ and $\Sigma ^{Z}(i\omega _{n})$. The first accounts for the
quasiparticle renormalizations which shifts the bare energies (real
part) and provides damping (imaginary part). The second accounts for
gap renormalization effects and also has frequency dependence and is
complex. We have evaluated numerically this complete equation for a
number of points in Fig.~\ref{fig3} and have verified that these
extra complications do not change in an important way, the
qualitative features shown in Fig.~\ref{fig3} which was drawn from
the evaluation of the simpler Eq.~(\ref{xy}). The numerical
differences between the two approaches agree to within a few percent
at $T\approx0K$ while at $T\approx200K$ they differ at the $20\%$
level. But this difference is not important for the conclusions we
will make. What is critical is the sign of the Hall conductivity
rather than its exact magnitude.

The boundary curves identified in Fig.~\ref{fig2} are reproduced on
Fig.~\ref{fig3} as long dashed black lines. Our finite temperature
data for the spin and valley Hall DC conductivity clearly reflects
some of the features of these sharp crossovers but temperature does
obscure their presence. Nevertheless our assignment of Chern numbers
to various phases treating the gap as if it were effectively a zero
temperature value, provides considerable insight into the origin of
the complex variation of spin and valley Hall conductivity seen in
Fig.~\ref{fig3}. The left frames include the electron-phonon
renormalization while the right frames do not and are for
comparison. The top frames give the spin Hall and the lower frames
give the valley Hall DC conductivity. We first note that in our
units, the spin Hall conductivity at T=0 is quantized and takes on
the value zero for the larger values of $|\Delta_{z}|$ and $-1/\pi$
in the region of smaller $|\Delta_{z}|$. With a discontinuous
crossover at $\tilde{\Delta}_{s_{z}}^{\tau }=0$. Similarly the
valley Hall conductivity jumps from $1/\pi$ to 0 and to $-1/\pi$
from right to left with boundaries also at
$\tilde{\Delta}_{s_{z}}^{\tau }=0$. As temperature is increased
slightly the left (with electron-phonon) and right (without
electron-phonon) panels remain practically indistinguishable. The
crossover is no longer a jump but occurs over a small region of
variation in $\Delta_{z}$. However, at higher temperatures they
become radically different. Without electron-phonon renormalization
the sign of the spin Hall conductivity remains unchanged, while
there can be sign changes when it is included. The region of small
$\Delta_{z}$ is particularly interesting as we go from a topological
insulator with effective Chern spin Hall number -1 to +1 by
increasing T. The spin Hall conductivity goes from a negative at
$T=0K$ to a positive value at $T=250K$ and this change in sign is
taken as an indication of a topological crossover. It occurs only
when the zero frequency value of the gap
$\tilde{\Delta}_{s_{z}}^{\tau }$ changes sign with increasing $T$.
This can never occur in a bare band picture and is a result of the
electron-phonon renormalizations which become larger with increasing
temperature and have the opposite sign to the bare gap $\Delta
_{s_{z}}^{\tau }=\Delta _{z}-\lambda _{SO}\tau s_{z}$. Note that for
a given valley the renormalized gap must be positive for one spin
state and negative for the other so that both contributions add in
the definition of the spin Hall conductivity to give a non zero
value. Without a sign change the two contributions would cancel
leading to zero spin Hall (trivial band insulator case).

In Fig.~\ref{fig4} we present results for the transverse AC
conductivity based on Eq.~(\ref{cxy}) with matrix spectral density
$\hat{A}$ obtained from the fully interacting Green's function
specified in Eq.~(\ref{full}) to (\ref{Green}). The left frame gives
the imaginary part of $\sigma_{xy}(\omega)$ in units of
$e^{2}/\hbar$ while the right gives the real part of the spin Hall
$\sigma^s_{xy}(\omega)$. Top to bottom frames give three
temperatures $T=10K$ (top), $T=160K$ (middle) and $T=250K$ (bottom).
In all cases the red curves are based on the bare bands for which
Eq.~(\ref{cxy}) simplifies to Eq.~(\ref{xy}). These are included for
comparison with the dotted curves which fully include the
electron-phonon renormalizations through the self energies $\Sigma
^{I}(i\omega _{n})$ and $\Sigma ^{Z}(i\omega _{n})$ of Eq.~(3).
Considering first the imaginary part $Im\sigma_{xy}(\omega)$ (left
frames) we note that at low temperature $T=10K$ bare (solid) and
renormalized (dotted) results are not qualitatively different from
each other. This is not the case at high temperatures. In particular
for $T=160K$ the sign associated with red ($\tau s_{z}=-1$) and
black ($\tau s_{z}=+1$) curves is no longer the same. This arises
because the sign of the renormalized results has reversed while that
of the bare band results is unaltered. This shows once more how
phonons can introduce temperature variation in the conductivity not
part of bare band calculations. For $T=250K$ the situation is even
more complex with renormalized conductivity crossing at
$\hbar\omega$ ~30meV (dotted curves). Turning next to the real part
of the spin Hall conductivity ($Re\sigma^s_{xy}(\omega)$) shown on
the right in Fig.~\ref{fig4} we note again that renormalized
(dotted) and bare band (solid) results agree qualitatively at low T
but that this changes as T increases. Particularly important for
this paper is the $\omega=0$ limit which gives the DC value of the
Hall conductivity used in Fig.~\ref{fig3}. At low temperature bare
and renormalized value agree while at higher T they in fact carry
opposite sign as has been emphasized in constructing the color plots
of Fig.~\ref{fig3}.

\section{Circular polarization and dichroism}

Another consequence of the phonon renormalization of the effective
gap is the change it can induce in the valley Hall optical selection
rule for circularly polarized light. In Fig.~\ref{fig5} we show
results for the frequency dependence of $Re\sigma _{\pm }(\omega )$
vs. $\omega$ in units of $\frac{e^{2}}{\hbar}$ at $T=30K$ (upper
panel) and at $T=300K$ (lower panel). The solid red curve is
$Re\sigma _{+}(\omega )$ for the spin up band and the red dotted is
for spin down. Both show significant absorption above the main
absorption edge determined by the effective low temperature gap. By
contrast the black curves give $Re\sigma _{-}(\omega )$ and show
that for this valley $\tau=1$, there is comparatively little
absorption of the left handed polarized light. This selection rule
can be reversed by increasing temperature through a change in the
sign of the effective gap brought about by the coupling to phonons
as seen in the lower frame for $T=300K$. This temperature is
sufficiently large and the gap small enough that no absorption
threshold can now be seen but it is clear that the left handed light
is now more strongly absorbed than the right handed light. Finally
we note sharp phonon structures at energies ($2|\Delta
_{s_{z}}^{\tau }|+2\omega_{E}$) at low temperature. Of course once
the sum over the two valleys $\tau=\pm1$ is carried out these
effects average out. However including a valley symmetry breaking
term to our Hamiltonian (1) would allow the effect to be seen
directly in optics.

\section{Conclusion}

Coupling of Dirac fermions to a phonon field can change the sign of
their effective gap with increasing temperature. This leads to a
rich set of crossovers from topological to band insulators with
recognizable imprints in their DC spin Hall and valley Hall
conductivity. It also leads to a switch with increasing temperature
in the polarization from right to left handedness of the dominantly
absorbed light by a given valley. Both effects provide a pathway to
the experimental verification of such a phonon induced topological
crossover.

\begin{acknowledgments}
This work was supported by the Natural Sciences and Engineering
Research Council of Canada (NSERC), the Canadian Institute for
Advanced Research (CIFAR).
\end{acknowledgments}

\end{document}